\documentclass{PoS}

\title{Some continuum physics results from the lattice V-A correlator}

\ShortTitle{Some continuum physics results from the lattice V-A correlator}

\author{P.A. Boyle,$^a$, L. Del Debbio,$^a$ N. Garron,$^b$, R.J. Hudspith,$^a$,
E. Kerrane$^c$, \speaker{K. Maltman},$^{d,e}$ and J.M. Zanotti$^e$ 
\\
\llap{$^a$}Physics and Astronomy, University of Edinburgh, Edinburgh
EH9 3JZ, UK\\
\llap{$^b$}School of Mathematics, Trinity College, Dublin 2, Ireland\\
\llap{$^c$}Instituto de F\`isica T\`eorica UAM/CSIC, Universidad
Aut\`onoma de Madrid, Cantoblanco E-28049 Madrid, Spain\\
\llap{$^d$}Mathematics and Statistics, York University, Toronto M3J 1P3
Canada\\
\llap{$^e$}CSSM, University of Adelaide, Adelaide 5005 Australia\\
E-mail: \email{paboyle@ph.ed.ac.uk},
\email{ldeldebb@ph.ed.ac.uk}, \email{ngarron@maths.tcd.ie},
\email{s0968574@sms.ed.ac.uk}, \email{eoin.kerrane@gmail.com},
\email{kmaltman@yorku.ca}, \email{james.zanotti@adelaide.edu.au}}

\abstract{We present preliminary results on 
extractions of the chiral LECs $L_{10}$ and $C_{87}$ and 
constraints on the excited pseudoscalar state
$\pi (1300)$ and $\pi (1800)$ decay constants obtained from an 
analysis of lattice data for the flavor $ud$ light quark
V-A correlator. A comparison of the results for the correlator 
to the corresponding mildly-model-dependent continuum results (based
primarily on experimental hadronic $\tau$ decay data) is also given.}

\FullConference{The 30th International Symposium on Lattice Field Theory\\
		 June 24-29,  2012\\
		 Cairns, Australia}
\begin{document}

\section{The V-A correlator}
We focus on the difference of flavor $ud$ vector (V) 
and axial vector (A) current-current 2-point functions, $\Pi_{V/A}^{\mu\nu}$, 
and their $J=0,1$ scalar components, $\Pi_{V/A}^{(J)}$, 
defined in Minkowski space by

\begin{eqnarray}
&&\Pi_{V/A}^{\mu\nu}(q^2)\, \equiv i\int\, d^4x\, e^{iq\cdot x}
\langle 0\vert T\left( J_{V/A}^\mu (x) J_{V/A}^{\dagger\, \nu} (0)\right)
\vert 0\rangle
\nonumber\\
&&\ \  
\,=\,
\left( q^\mu q^\nu - q^2 g^{\mu\nu}\right)\, \Pi^{(1)}_{V/A}(Q^2)
\, +\, q^\mu q^\nu\, \Pi_{V/A}^{(0)}(Q^2)\, 
\label{mink2pt}\end{eqnarray}
where, as usual, $Q^2\, =\, -q^2$. In what follows, we denote 
$\Delta\Pi^{(J)}\equiv \Pi_V^{(J)}-\Pi_A^{(J)}$.
The $\Pi_{V/A}^{(J)}(Q^2)$, for $Q^2>0$, also determine the corresponding
Euclidean 2-point functions 
\begin{eqnarray}
&&\left[\Pi_{V/A}^{\mu\nu}(Q^2)\right]_{Eucl}\, =\,
\left( Q^2\delta^{\mu\nu}- Q^\mu Q^\nu\right)\, \Pi^{(1)}_{V/A}(Q^2)
\, -\, Q^\mu Q^\nu\, \Pi_{V/A}^{(0)}(Q^2)\, ,
\label{eucl2pt}\end{eqnarray}
making the $\Pi_{V/A}^{(J)}(Q^2)$ accessible from lattice simulations.
$\Pi_{A}^{(0)}$ and $\Pi_{A}^{(1)}$ both have kinematic poles at
$Q^2=0$ while the $J=0+1$ sum does not. Since, beyond NLO in the
chiral expansion, the pole residues involve
at-present-unknown NNLO LECs, we focus on 
%\begin{equation}
$\Delta\Pi (Q^2)\equiv \Pi_V^{(0+1)}(Q^2)-\Pi_A^{(0+1)}(Q^2)$,
%\label{vmadefn}\end{equation}
which satisfies an unsubtracted dispersion relation with only
physical singularities. The corresponding spectral function, $\Delta\rho (s)$,
consists of a $\delta$-function at $s=m_\pi^2$ and continuum beginning at 
$4m_\pi^2$. The ``continuum part'', $\Delta\overline{\Pi}(Q^2)$, of
$\Delta\Pi (Q^2)$ results from removing the $\pi$ pole:
$\Delta\Pi (Q^2)\, =\, \Delta\overline{\Pi}(Q^2)\, -\, 2f_\pi^2/(Q^2+m_\pi^2)$.
For $s<m_\tau^2$, $\Delta\rho (s)$ can be determined experimentally from 
hadronic $\tau$ decay data~\cite{taubasics}. Public versions are
available from both ALEPH~\cite{alephvma} and OPAL~\cite{opalvma},
with a yet-to-be-corrected problem affecting the covariance matrix
of the former. Beyond $s=m_\tau^2$, 
an alternate representation results from fitting a
physically motivated model for duality violations (DVs)~\cite{cgp08} 
to integrated versions of the data~\cite{dv7}. The OPAL data and 
fitted DV model provide a dispersive determination of $\Delta\Pi (Q^2)$ 
at spacelike $Q^2>0$, where it can also be measured on the lattice.
The dispersive result is nominally quite precise, but 
has some (mild) model-dependence from the use of the DV model.
It also involves $\tau\rightarrow 4\pi\nu_\tau$ 
contributions in the $V$ channel whose uncertainties 
may have been underestimated, given that the corresponding
$4\pi$ branching fractions differ from expectations based on
CVC and measured $e^+e^-\rightarrow 4\pi$ cross-sections by much more 
than is typical for isospin-breaking corrections~\cite{dehz4pi,new4pielectro}.
The mildness of the model-dependence follows from the fact that,
in the range $Q^2< (500\ MeV)^2$ expected to be of
relevance to the determination of chiral LECs, DV contributions
to $\Delta\rho (s)$ account for a few to several $\%$
of $\Delta\overline{\Pi} (Q^2)$, the precise values depending on the point
in the spectrum at which one switches from data to the
fitted DV model. The cross-check on the lattice results for
$\Delta\overline{\Pi}$ provided by the dispersive representation is also
useful in light of the freedom to vary $m_{u,d,s}$ on the lattice, which, 
in principle, provides access to NNLO chiral LECs currently unknown and/or 
difficult to extract reliably with continuum methods.

\section{Lattice data for $\Delta\Pi (Q^2)$}
\begin{figure} [htb]
%\unitlength1cm
%\center{\begin{minipage}[t]{10.0cm}
%\begin{picture}(9.9,9.9)
\centering
\rotatebox{270}{\mbox{
\includegraphics[width=0.65\textwidth]
{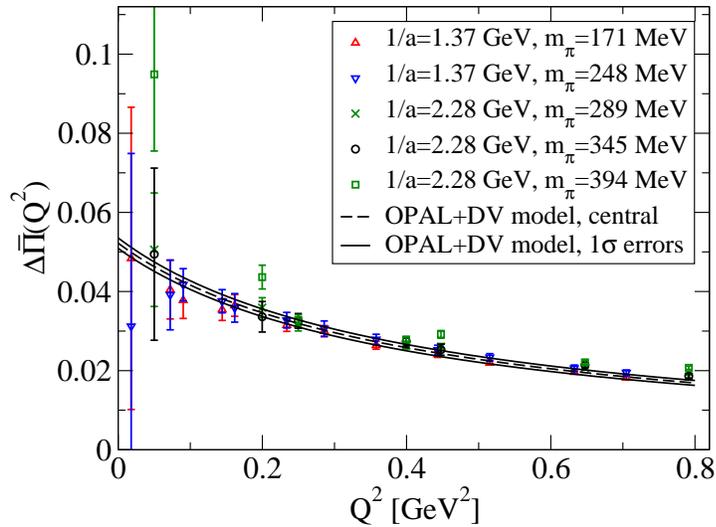}
%\end{picture}
%\end{minipage}
}}
\caption{Lattice and OPAL+DV model results for $\Delta\overline{\Pi} (Q^2)$}
\label{fig1}
\end{figure}

$\Delta\Pi (Q^2)$ has been determined for the fine $1/a=2.28$ GeV,
$m_\pi = 289$, $345$ and $394$ MeV, and coarse $1/a=1.37$ GeV,
$m_\pi = 171$ and $248$ RBC/UKQCD DWF ensembles detailed in 
Refs.~\cite{ainv228,ainv137}. The latter provide an increased 
number of low-$Q^2$ points, improving the 
determination of the chiral LECs. The values of $f_\pi$ and $m_\pi$ 
needed to convert $\Delta\Pi (Q^2)$ to $\Delta\overline{\Pi}(Q^2)$ 
are given in Refs.~\cite{ainv228,ainv137}. The simulation $m_s$ values 
are, in all cases, close, but not exactly equal to the physical value. 
As the $SU(2)$ LECs correspond to physical $m_s$, we analyze the data in the 
chiral $SU(3)$ framework. The ensemble $m_K$ values required for
this purpose are also given in Refs.~\cite{ainv228,ainv137}. The spectral
function, $\Delta\overline{\rho}$, of $\Delta\overline{\Pi}$, and hence
also $\Delta\overline{\Pi}$ itself, are $O(m_\ell^0)$ in the chiral expansion.
We thus expect the lattice $\Delta\overline{\Pi}(Q^2)$ to approach the
physical results for sufficiently light $m_\ell$. We find that,
within errors, for the low-$Q^2$ region of interest to us here,
the lattice $\Delta\overline{\Pi}(Q^2)$ agree well with one another,
and with the continuum OPAL+DV model results, 
for all but the $m_\pi =394$ MeV case, as shown in Fig.~\ref{fig1}. 

\section{The chiral LECs}
\begin{figure} [htb]
%\unitlength1cm
%\center{\begin{minipage}[t]{10.0cm}
%\begin{picture}(9.9,9.9)
\centering
\rotatebox{270}{\mbox{
\includegraphics[width=0.65\textwidth]
{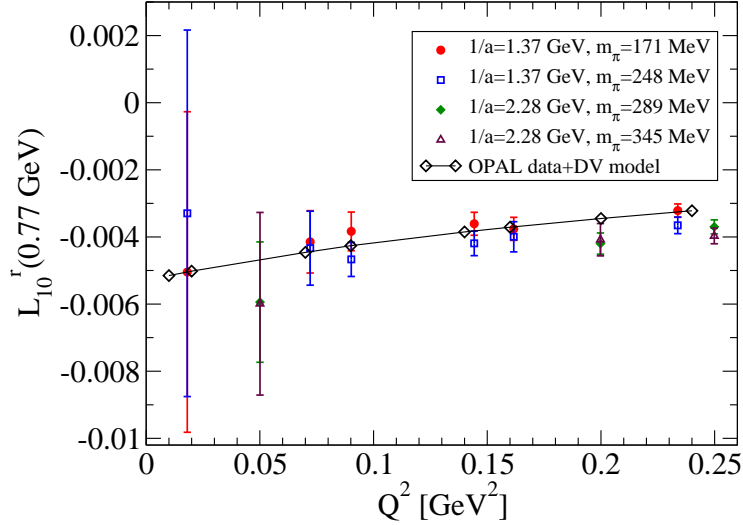}
%\end{picture}
%\end{minipage}
}}
\caption{NLO results for $L_{10}^r(\mu_0 )$ from analyses of
lattice and OPAL+DV model versions of $\Delta\overline{\Pi}$ }
\label{fig2}
\end{figure}

At NLO in the chiral expansion, $\Delta\Pi$ is 
controlled by the single NLO LEC $L_{10}^r(\mu )$. Two previous lattice 
studies determined $L_{10}^r(\mu )$ by analyzing 
$\Delta\Pi^{(1)}$~\cite{jlqcdvma,rbcukqcdvma09} at NLO. With
the lattice spacings available, the second-smallest non-zero 
$Q^2$ were found to be too large ($\sim (650\ MeV)^2$ and $\sim (460 MeV)^2$ 
for Refs.~\cite{jlqcdvma} and ~\cite{rbcukqcdvma09} respectively) to
allow for a successful NLO fit. Final NLO analysis results were thus based on
the single lowest $Q^2$ values, $(320\ MeV)^2$ and $\sim (230\  MeV)^2$,
respectively. The current analysis improves on the previous ones in a 
number of ways. First, the new coarser lattices allows access to an 
increased number of low-$Q^2$ points. Second, the statistics for
the $m_\pi =289$ MeV ensemble considered previously~\cite{rbcukqcdvma09} 
have now been doubled. Third,
because the residue of the $Q^2=0$ kinematic pole in $\Delta\Pi^{(1)}$
involves an unknown NNLO contribution, whose contribution, 
relative to that of the term involving the NLO constant $L_{10}^r$, 
gets enhanced when one goes to the low $Q^2$ desirable for extracting LECs, 
we switch to analyzing $\Delta\overline{\Pi}$ rather than $\Delta\Pi^{(1)}$,
the $\pi$ pole contribution which must be subtracted to obtain
the former having both an exactly known residue and being farther
from the region of the lattice data than is the kinematic pole 
in $\Delta\Pi^{(1)}$.
The NLO results for $L_{10}^r(\mu )$, for $\mu =\mu_0 =0.77\ GeV$, obtained 
for each $Q^2<0.25\ GeV^2$, and all but the heaviest $m_\pi =394
\ MeV$ ensemble, are shown in Fig.~\ref{fig2}. Also shown for
comparison are the results of an NLO analysis of the OPAL+DV model 
over a similar range of $Q^2$. 

Since the $Q^2$-dependence of $\Pi_V(Q^2)$ is known to be poorly 
reproduced by the NLO representation~\cite{ab06}, one might be 
surprised by the relative stability of the results for different $Q^2$. 
However, the missing intermediate $\rho$ contribution believed responsible 
for the NLO $\Pi_V$ slope problem~\cite{ab06} is encoded in the NNLO 
LEC $C_{93}^r$~\cite{abt00}, and the contribution to $\Pi_V(Q^2)$ proportional
to $C_{93}^r$ is exactly cancelled by that proportional to $C_{93}^r$ 
in the NNLO expression for $\Pi_A^{(0+1)}(Q^2)$.
Nonetheless, the central value for the average slope
of $\Delta\overline{\Pi}(Q^2)$ with respect to $Q^2$ is significantly
larger than expected from the NLO expression, albeit at only the
$\sim 2\sigma$ level. The structure of the full NNLO result, 
known from the results of Ref.~\cite{abt00}, is very linear in
$Q^2$ for the $Q^2$ considered here, so a significant portion of the NNLO 
contribution is easily removed by fitting the results of Fig.~\ref{fig2} 
to a linear form and using this to extrapolate to $Q^2=0$. 
The only NNLO contributions remaining to be removed are then
those entering $\Delta\overline{\Pi}(0)$. These involve two NNLO LEC combinations,
$C_{61}^r -C_{12}^r-C_{80}^r$, which is not large-$N_c$ suppressed,
and $C_{62}^r -C_{13}^r-C_{81}^r$, which is~\cite{abt00,gappvma08}.
The first combination has been estimated  in Ref.~\cite{gappvma08}
using the results of previous continuum works. The second is
currently unknown, and has only been loosely bounded,
using rough large-$N_c$-suppression arguments~\cite{gappvma08}.
%(See Ref.~\cite{gappvma08} for references to the 
%relevant earlier works, and further details.) 
The resulting NNLO LEC combination assessments were used in 
Ref.~\cite{gappvma08} to obtain a continuum extraction of $L_{10}^r$ and 
$C_{87}^r$ (the NNLO LEC expected to dominate the slope of $\Delta\overline{\Pi}$).
The analysis employed the ALEPH data and was based on two 
additional, not explicitly tested, assumptions, namely 
(i) that the NNLO form will successfully represent 
$\Delta\overline{\Pi}(Q^2)$ and (ii) that the V and 
A channel DV spectral contributions which, being governed by 
the resonance structure in the channel in question, are expected to be 
channel-dependent, can be assumed to be approximately the same in
form and hence combined into a single DV ansatz for the V-A difference.
While the latter assumption is not borne out by the combined V and A channel
fits of Refs.~\cite{dv7}, the contribution to $\Delta\overline{\Pi}$
of the DV part of $\Delta\overline{\rho}$ is small in the low-$Q^2$ region.
It is thus of interest to compare the results of our fit to those
of this mildly model-dependent continuum analysis, which are~\cite{gappvma08}
\begin{equation}
L_{10}^r(\mu_0 )\, =\, -0.0041(4)_{NNLO};
\qquad C_{87}^r(\mu_0 )\, =\, 0.0049(2)_{NNLO}\ GeV^{-2}\, ,
\label{lecresults}\end{equation}
with the error dominated entirely by the uncertainty in the estimate
for unknown large-$N_c$-suppressed NNLO LEC combination. 
Results obtained using instead the OPAL data, 
and incorporating the results of the DV model fits of Refs.~\cite{dv7} in 
order to remove the second of the two assumptions noted above, are in
extremely close agreement. An additional unknown systematic error, associated
with the two additional assumptions noted above, of course also exists
for the results of Eq.~(\ref{lecresults}). Our final goal is to perform a 
NNLO analysis of the lattice data, including a range of $m_q$ sufficient
to put constraints on the currently unknown NNLO LECs (something
likely to be feasible given the difference of the results for
$\Delta\overline{\Pi}$ for $m_\pi =394\ MeV$ from those for the
smaller $m_\pi$ seen in Fig.~\ref{fig1}), but this analysis
has not yet been completed. The results of following
the continuum estimates/arguments of Ref.~\cite{gappvma08} 
for the unknown NNLO LEC combinations entering $\Delta\overline{\Pi}(0)$ are
\begin{equation}
L_{10}^r(\mu_0 )\, =\, -0.0038(4)_{latt}(4)_{NNLO};
\qquad C_{87}^r(\mu_0 )\, =\, 0.0040(21)_{latt}(2)_{NNLO}\ GeV^{-2}\, ,
\label{lecresultslatt}\end{equation}
in good agreement with, though less precise than the NNLO-LEC-induced
part of the error obtained explicitly in the continuum analysis of
Ref.~\cite{gappvma08}.

\section{Constraints on the $\pi^\prime$ and $\pi^{\prime\prime}$
Decay Constants}

Excited $I=1$ pseudoscalar mesons, $P$, couple with strengths
$2f_{P}m_{P}^2$ to the divergence of the flavor $ud$ axial current. Their
decay constants, $f_P$, enter the conventional determination 
of $m_u+m_d$ employing sum rules for the two-point function of 
this divergence~\cite{sumrulemhat} and are currently determined as 
part of the analysis. A similar internal determination is 
required in the extraction of $m_u+m_s$ 
from sum rules for the two-point function of the divergence of the flavor 
$us$ axial current. Finally, flavor-breaking sum rules used to
determine $\vert V_{us}\vert$ from hadronic $\tau$ 
decay~\cite{gamizetal,kmcw}, or a combination of hadronic $\tau$ decay 
and electroproduction cross-section data~\cite{kmtauem}, 
encounter a problem with the very bad convergence of the 
OPE representation of $J=0$ contributions, necessitating the
subtraction of the chirally suppressed, but not totally negligible,
strange excited state scalar and pseudoscalar contributions to the 
differential $\tau$ decay spectrum. The strange pseudoscalar subtraction 
relies on the excited $K$ decay constants obtained in the sum rule analysis. 

\begin{figure} [htb]
%\unitlength1cm
%\center{\begin{minipage}[t]{10.0cm}
%\begin{picture}(9.9,9.9)
\centering
\rotatebox{270}{\mbox{
\includegraphics[width=0.6\textwidth]{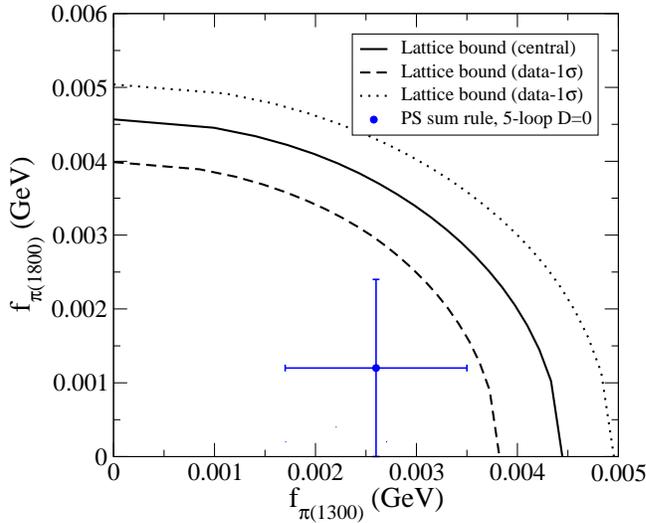}
%\end{picture}
%\end{minipage}
}}
\caption{Lattice constraints on the $\pi^\prime$ and $\pi^{\prime\prime}$
decay constants}
\label{fig3}
\end{figure}

The lattice data allows us to test the reliability of the sum rule 
determination of such decay constants, as follows. The
quantity $P(Q^2)\, \equiv Q^2\, \Delta\Pi^{(0)}(Q^2)$, which,
for $m_u=m_d$ is equal to $ -Q^2\, \Pi^{(0)}_{A}(Q^2)$, is free of
kinematic singularities and satisfies a once-subtracted dispersion
relation. Since $\Pi^{(0)}_A(Q^2)$ and the quantities $m_\pi$,
$f_\pi$ which determine the pion pole contribution to the dispersive
representation are all measurable on the lattice, the following
rearranged version of this relation provides constraints on the
continuum contribution, and hence on the excited state decay
constants, $f_{\pi^\prime}$ and $f_{\pi^{\prime\prime}}$, for each
pair of $Q^2$ and subtraction point $Q_0^2$:
\begin{eqnarray}
&&P(Q^2)\, -\, P(Q_0^2)\, +\, {\frac{(Q^2-Q_0^2)\, 2f_\pi^2 m_\pi^2}
{(s+Q^2)(s+Q_0^2)}}\, =\, 
%\nonumber\\
%&&\qquad\quad 
-(Q^2-Q_0^2)\int_{9m_\pi^2}^\infty ds\,
{\frac{s\, \rho^{(0)}_A(s)}{(s+Q^2)(s+Q_0^2)}}\, .
\label{expsconstraint}\end{eqnarray}
Spectral positivity ensures that the LHS provides an upper bound on the
contributions from any subset of the full set of excited pseudoscalar
states. In the narrow width approximation, this constraint represents
a straight line in the $f^2_{\pi^\prime}$-$f^2_{\pi^{\prime\prime}}$
plane for each pair $(Q^2,Q_0^2)$. The fact that the excited state
decay constants scale as $m_\pi^2$ can be used to scale each such
bound from the masses used in the simulation down to physical $m_\pi$.
It turns out that, at present, 
only the high-statistics, $1/a=2.28\ GeV$, $m_\pi =289\ MeV$ 
ensemble provides data sufficiently accurate for this purpose.
The envelope of the resulting set of  
constraint lines, scaled down to physical $m_\pi$, is shown in 
Fig.~\ref{fig3}. Also shown, for comparison, are the results obtained/used
in Refs.~\cite{sumrulemhat}. These are obviously in good agreement
with the lattice constraints, leaving room for small additional contributions
to the dispersive representation from yet higher excited pseudoscalar
resonances.

\section{Acknowledgements}
The computations were done using the STFC's DiRAC facilities at Swansea
and Edinburgh. PAB, LDD, NG and RJH are supported by an STFC Consolidated 
Grant, and by the EU under Grant Agreement PITN-GA-2009-238353 (ITN
STRONGnet).
EK was supported by the Comunidad Aut\`onoma de Madrid under
the program HEPHACOS S2009/ESP-1473 and the European Union under Grant 
Agreement PITN-GA-2009-238353 (ITN STRONGnet).
KM acknowledges the hospitality of the CSSM, University of Adelaide,
and support of NSERC (Canada). 
JMZ is supported by the Australian Research Council grant FT100100005.

\end{document}